# CloakLM: Obfuscating GPU Memory Layout to Mitigate Model Ex-filtration for Serving

Kunal Jain
Georgia Institute of Technology
USA
kunal.jain@gatech.edu

Seokjin Go
Georgia Institute of Technology
USA
seokjin.go@gatech.edu

Divya Mahajan
Georgia Institute of Technology
USA
divya.mahajan@gatech.edu

## Abstract

Large foundation models deployed on third-party and shared accelerator infrastructure face a practical risk of model exfiltration that existing defenses do not fully address [30]. In common serving deployments, the model provider controls the VM or bare-metal serving stack and trusts the drivers and runtime of its own accelerators, but does not control the surrounding hardware substrate: the host–GPU interconnect, the accelerator fabric, and neighboring infrastructure components remain outside the tenant's trust boundary. Prior work has demonstrated that both attack surfaces are practically exploitable. Hermes [39] shows that a passive observer with access to the PCIe bus can achieve lossless DNN reconstruction from transferred packets alone. TunnelS [36] shows that a non-participating host with driver-level access can exfiltrate HBM contents at high throughput by exploiting PCIe underutilization, without interrupting inference. Beyond these direct hardware paths, co-tenant VMs reachable over the shared frontend or management network can access memory-mapped interfaces or improperly segmented RDMA regions without requiring physical co-location.

These attacks share a common enabler: ML frameworks allocate model weights in large, contiguous, and repeatedly accessed memory regions to maximize throughput. This regularity makes intercepted PCIe transfers and raw HBM captures structurally rich enough to reconstruct layer boundaries, tensor shapes, and model parameters.

We present CloakLM, a software-only memory-obfuscation framework that removes this structural regularity without modifying the inference stack's logical view of memory. CloakLM operates across three tiers: PCIe traffic shaping to reduce extraction bandwidth on the host–GPU path; inter- and intra-layer weight shuffling to break the correspondence between observed access sequences and model structure; and physical HBM page remapping to ensure raw memory captures no longer yield contiguous, reconstructable model segments. The authorized execution sees a valid virtual memory layout and incurs negligible overhead; whereas unauthorized observers see fragmented, semantically incoherent state.

CloakLM integrates with vLLM and PyTorch, requires no hardware changes, and is complementary to confidential computing. We evaluate CloakLM on distributed inference workloads using LLaMA and Qwen models and show near-native performance alongside strong resistance to PCIe snooping and HBM dump attacks, demonstrating that inference-time model exfiltration can be made substantially less practical in real-world deployments.

## 1 Introduction

Large foundation models [10, 20, 26] are increasingly deployed as remote services on accelerators hosted in third-party datacenters and shared GPU clusters [19]. Providers such as OpenAI [26], Anthropic [1], and Mistral [15, 16] relinquish physical control of the hardware, yet the deployed model represents their primary intellectual property and economic asset. While the community has extensively studied secure VMs, data isolation, and user-data confidentiality, the security of *model weights* as they traverse and reside in accelerator memory remains largely unaddressed.

**Trust boundaries in third-party inference deployments.** In practice, a model provider is granted control over a virtualized host or VM instance provisioned by the cloud operator. We treat this VM and its user-space serving stack as trusted, it executes the provider's own software and is comparable to bare-metal access granted by a CSP. The trust boundary lies below and around this VM along two dimensions. On the *software side*, we trust the accelerator drivers, firmware, and runtime stack on GPUs that are explicitly part of the serving deployment, these are provisioned to the model owner and execute software under their control. On the *hardware side*, the physical accelerator fabric, interconnects, peer hosts, and nodes on the same front-end network are *not* under the model provider's control: an attacker need not subvert any trusted software component to reach model weights if they have access to the underlying hardware substrate or the network plane connecting it.

As shown in Figure 1, this boundary has material consequences as a model transitions from training to inference. During training, parameters are short-lived, continuously evolving, and confined to a small number of tightly managed clusters; this transience lets providers exercise close operational control over where and how parameters reside. Instead, at inference time stable checkpoints are replicated, persistently resident on third-party and disaggregated accelerators, and kept resident for long periods while serving requests. Even when the serving VM is fully trusted, *any actor with access to the same hardware substrate, accelerator HBM, or co-resident VM can potentially observe, access, or extract model weights.*

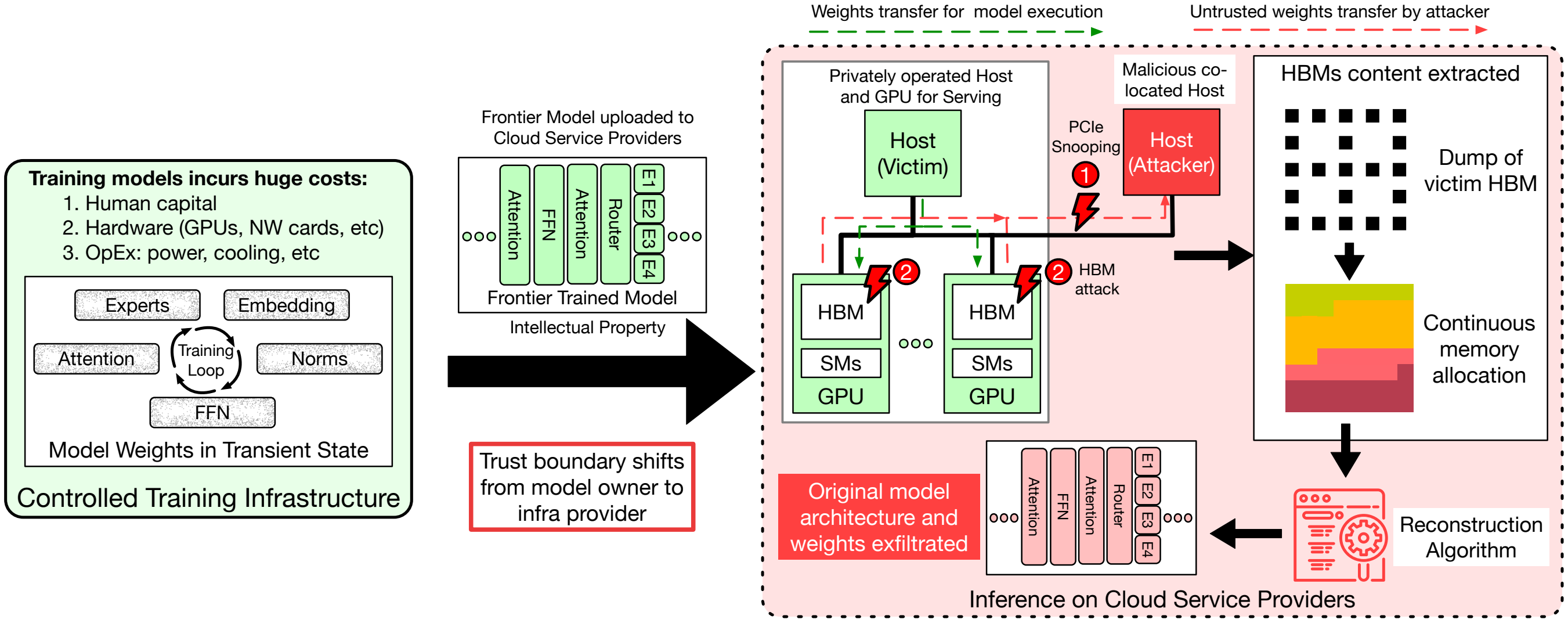

**Figure 1. Model lifecycle and inference-time memory exposure.** During training, model parameters are transient and reside on tightly controlled infrastructure. Once deployed for inference, stable checkpoints cross the trust boundary and become persistently resident on third-party, disaggregated accelerators. Co-tenant VMs, non-participating GPUs on the same fabric, and physical bus observers can all access or intercept model weights, through direct memory channels, shared interconnects, or PCIe snooping, without participating in or disrupting the inference workload.

**A concrete and expanding attack surface.** We identify three classes of inference-time model exfiltration attacks, each realistic in today's cloud and datacenter deployments. The first targets *host- and VM-level access*: a malicious co-tenant or non-participating host can reach HBM over the shared management network, or exploit idle PCIe bandwidth through driver-level modifications, without touching the serving stack [36]. The second targets the *accelerator fabric*: a non-participating GPU on the same NVLink domain or interconnect fabric can access peer device memory or intercept its traffic at the device level, without requiring host-side code execution or network reachability [25]. The third involves *physical hardware access*: an attacker with physical proximity to the server can instrument the infrastructure directly, for example by inserting a PCIe snooping device, and reconstruct model parameters from DMA traffic or device memory without involving the host software stack [39]. All three positions share the same capability and limitation: they obtain *raw memory contents or raw bus transfers* outside the serving stack, without the framework-level metadata such as virtual memory layout, virtual-to-physical page mappings, and tensor placement records.

**Why ML software stacks make raw memory access sufficient.** Even though the attacker does not have framework-level metadata, in an unprotected deployment this limitation is immaterial. Frameworks such as PyTorch and vLLM allocate model weights in large, *contiguous* physical regions that are *persistently resident* and *repeatedly accessed* in a predictable order to maximize throughput. This regularity makes both memory contents and bus-visible transfers semantically rich: contiguity, stable access order, and tensor regularity are alone sufficient to infer layer boundaries, tensor shapes, and parameter values, the framework metadata is effectively recoverable from the raw data itself, without access to the serving stack [31, 36, 39].

**Protecting model-resident memory beyond confidential computing.** Confidential Computing (CC) provides strong hardware-backed isolation when the entire execution environment supports it, and CloakLM is designed to complement it. CC requires the trust boundary to extend across every participating component: a single non-enclave node or heterogeneous accelerator breaks the chain, and protects HBM through access controls rather than encrypting memory contents at runtime [11]. Physical access to the device through hardware probing or firmware-level bypass can therefore still recover plaintext weights directly from HBM, even on CC-enabled hardware. More broadly, CC addresses data in motion across the protected fabric, but does not reduce an attacker's ability to reconstruct the model from weights already resident in device memory. CloakLM targets this complementary surface: by fragmenting and permuting the physical layout of model weights in HBM, raw memory captures, however obtained, yield structurally incoherent data rather than reconstructable model parameters, on any infrastructure regardless of enclave support.

**CloakLM: obfuscation as an accelerator-memory defense.** CloakLM reshapes how model weights are laid out and transferred across the accelerator memory path. Tier 0 applies controlled PCIe traffic shaping to reduce extraction bandwidth and fidelity on the host–GPU path. Tier 1 randomizes layer placement so that observed access sequences no longer reveal execution order or contiguous tensor regions. Tier 2 remaps the physical HBM pages backing model weights so that raw memory captures no longer yield contiguous, semantically coherent segments.

CloakLM is a software-only solution integrated with PyTorch allocators, the CUDA runtime, and vLLM, while preserving compatibility with distributed inference. Authorized execution retains a valid logical memory view and near-native performance; unauthorized observers who obtain raw memory dumps or bus-level captures see fragmented, incoherent state that resists reconstruction without the serving stack's layout metadata. CloakLM is designed specifically against attackers who access raw physical memory or bus transfers from outside the serving environment. We do *not* defend against full compromise of the host OS, hypervisor, or firmware of the serving stack, nor against attackers who can directly read virtual-to-physical mappings or modify inference kernels, such adversaries can defeat any software-only layout defense. Our goal is *cost imposition, not secrecy*: making accelerator-side model extraction substantially less practical for the attacker classes described above.

**Contributions.** We make the following contributions:

- **Defining accelerator-side attack vectors for model exfiltration.** We identify and characterize three concrete attacker positions, co-tenant or non-participating hosts VMs via the management network, non-participating accelerators via shared backend network fabric, and physical bus access, and show why existing defenses such as confidential computing and network isolation are either limited in deployment or do not practically fully address them.
- **Characterizing inference-time model exposure.** We show how contiguous virtual-to-physical placement and stable HBM residency in current serving stacks make raw memory access, without framework metadata, sufficient for practical model reconstruction.
- **A multi-tier software defense against accelerator-side exfiltration.** CloakLM disrupts extraction signals across PCIe-visible transfers, runtime-visible layout, and physical HBM placement by fragmenting, shuffling, and remapping model weights, without hardware changes or secure enclaves.
- **Performance-preserving design for memory-bound workloads.** Authorized execution retains a valid logical memory view and efficient GPU access behavior despite obfuscated physical placement, with no kernel modifications required.
- **End-to-end implementation and evaluation on real inference stacks.** CloakLM integrates with PyTorch and vLLM [17], supporting distributed multi-GPU execution, and is evaluated on LLaMA and Qwen, demonstrating strong extraction degradation at near-native inference performance.

## 2 Background

### 2.1 AI Infrastructure and Model Ownership

**Model weights are critical intellectual property.** Foundation models represent the primary intellectual property of modern AI companies [1, 16, 26]. Training a state-of-the-art model requires months of engineering effort, curated datasets at scale, and compute budgets measured in millions of dollars. The resulting weights encode this accumulated investment in a compact, executable form. For many providers, the trained model is the entire product: once ex-filtrated, it can be reused, fine-tuned, or served by a competitor at negligible marginal cost, eliminating the original provider's competitive advantage entirely. Protecting model weights is therefore a core business and national security requirement, not merely a privacy concern [30].

**This work focuses on the inference.** While protecting model weights across the full deployment lifecycle is important, training and inference present different exposure profiles that call for different approaches. Training workloads tend to run on more homogeneous, purpose-built clusters where infrastructure configuration is more tightly controlled. NVIDIA GPUs currently dominate this space, accounting for approximately 92% of the accelerator market [3]. Inference, by contrast, must scale across heterogeneous environments, rented accelerators, shared datacenters, and geographically distributed deployments and operates on stable, long-lived model checkpoints that remain unchanged across millions of requests. These two properties, the breadth of heterogeneous deployment and the stability of inference checkpoints, make inference both a high-risk and the most tractable phase for software-based weight protection. We focus on inference throughout this paper and leave training-time defenses to future work.

### 2.2 Inference Deployment and the Trust Boundary

**Inference is deployed on heterogeneous, third-party infrastructure.** To meet inference demand, providers rely on Cloud Service Providers (CSPs) such as AWS, Azure, and Google Cloud, renting accelerators hosted in third-party datacenters. The provider retains control of the software stack but cedes physical control of the hardware: the underlying servers, PCIe fabric, and accelerator interconnects are owned and operated by the CSP. Co-located tenants share network segments, accelerator racks, and in some configurations, the same physical hosts. This deployment model exposes model weights to infrastructure operators and co-located adversaries whose presence and capabilities the model provider cannot audit or control.

**Inference checkpoints are stable and persistently resident.** Unlike training, where parameters are updated continuously and intermediate states are transient, inference operates on fixed model checkpoints that remain resident in accelerator memory for extended periods. A single deployment may serve millions of requests with the same weights continuously loaded in GPU HBM. This stability creates a persistent, high-value target: an adversary with access to the accelerator or its interconnects can observe the same data repeatedly, aggregate observations over time, and correlate partial captures to reconstruct full model state.

### 2.3 System and GPU Architecture for Serving

**Weights are loaded once and persistently resident.** LLM inference is GPU-centric: model weights are transferred from host DRAM to GPU HBM at service startup and remain resident throughout the serving lifetime. Inference kernels read weights from HBM on every forward pass, but the parameters themselves are never modified. HBM therefore functions not as transient working memory but as a stable, read-only store of model parameters, continuously present in device memory for the duration of the deployment.

**Frameworks allocate parameters in large contiguous regions.** Inference frameworks allocate model parameters using bulk allocators optimized for throughput. PyTorch's caching allocator reserves large contiguous virtual memory regions and maps them to physical HBM pages in a size-ordered, predictable fashion [5]. vLLM's block allocator similarly assigns contiguous physical blocks to layer weight tensors to maximize memory bandwidth utilization during kernel execution [17]. As a result, a model's weight tensors occupy large, contiguous, physically predictable regions of HBM, laid out in the order in which they are loaded and subsequently accessed.

**Inference computation spans shared interconnects.** Modern inference deployments disaggregate computation across accelerators connected by high-speed fabrics. Within a host, GPUs communicate with the CPU over PCIe; across GPUs within a rack, NVLink provides high-bandwidth device-to-device transfers; across racks, RDMA-capable networks carry model shards and activations between nodes [4, 24]. These interconnects are shared infrastructure: in cloud deployments, the physical fabric and switching equipment are operated by the CSP, and multiple tenants' workloads may traverse the same switches and fabric segments.

### 2.4 Confidential Computing

**CC establishes a hardware-enforced trust perimeter.** Confidential computing (CC) platforms, such as NVIDIA Hopper and Blackwell Confidential Compute [11], address inference-time exposure by establishing a hardware-enforced enclave around CPU and GPU execution. Within this perimeter, PCIe transfers between host and GPU (data and kernel information) are encrypted in transit, and GPU kernels execute inside an enclave that excludes the hypervisor and CSP from the trusted computing base. NVLink connections between GPUs within the same CC domain are similarly encrypted and authenticated.

**HBM contents are access-controlled but not encrypted at rest.** CC enforces access control over HBM, preventing unauthorized code from reading device memory, but does not encrypt the underlying DRAM cells. Weight values reside in HBM as plaintext; the access-control layer governs who may read them, not whether the bits themselves are protected at the storage level. This is an architectural consequence of the GPU compute model: kernels require low-latency, high-bandwidth access to weights on every forward pass, and encrypting HBM at rest would impose prohibitive overhead on every memory access.

## 3 Threats for Model Exfiltration

Production inference deployments expose model weights through multiple simultaneous attack surfaces. Weights transferred to GPU HBM at service startup remain persistently resident and are repeatedly accessed in predictable patterns throughout the serving lifetime, making them an unusually stable target. The interconnect transfers that carry these weights between host and GPU are unencrypted and structurally regular, reflecting the deterministic access patterns of inference kernels. The shared management and frontend networks that connect co-located infrastructure provide additional reachability to the accelerator and its memory. Together, these properties place model weights in simultaneous contact with three distinct attacker positions, each operating at a different layer, each requiring a different kind of defense.

### 3.1 Attack Surfaces

In this work, the trust boundary runs at the serving software stack: the host OS, drivers, and runtime are trusted; the physical hardware and all non-serving infrastructure are not. Recent work has shown that accelerators, interconnects, and co-tenant VMs are increasingly exploitable attack surfaces. Figure 2 illustrates all three attacker positions relative to the trusted serving environment.

(1) **Co-tenant or non-participating host.** A compromised VM or non-participating host on the same infrastructure can reach model weights through two complementary paths. Over the shared front-end or management network, a co-tenant with authenticated access can exploit reachability to the serving host directly: the XZ Utils backdoor [13] illustrates this entry point, where a single compromised package granted network access to otherwise isolated hosts, and prior cross-tenant work [32, 33] has shown that network adjacency alone is sufficient to leak program state from victim tenants. At the driver level, a non-participating host need not use the management network at all: TunnelS [36] demonstrates that driver-level modifications can exfiltrate HBM contents at high throughput by exploiting idle PCIe windows, without executing any code on the victim host or interrupting its workload.

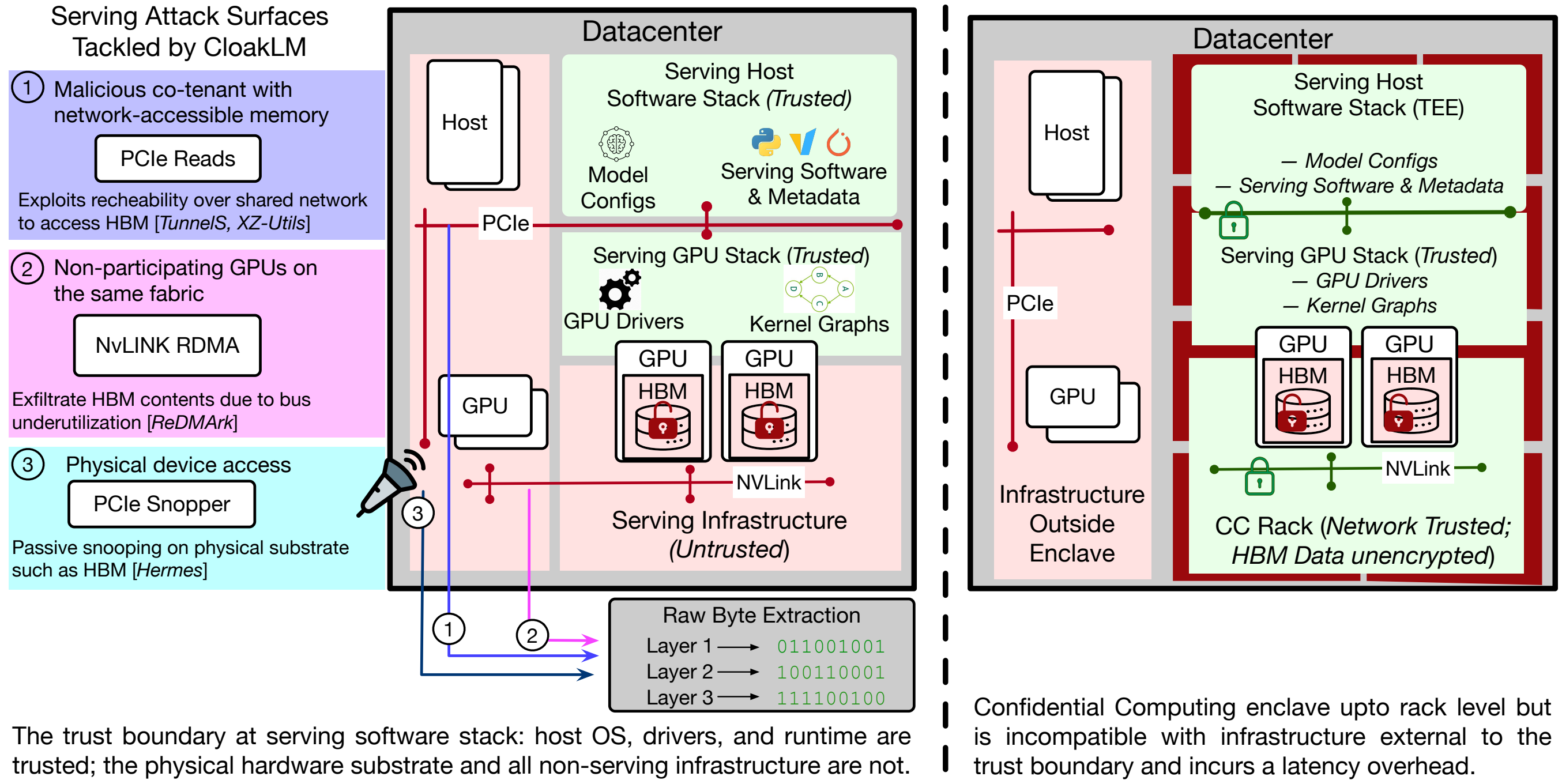

**Figure 2.** We evaluate three attack surfaces: ① A malicious co-tenant or non participating hosts, ② A malicious GPU on the same accelerator fabric and, ③ An adversary with physical access to the infrastructure. The adversary in all three scenarios is able to extract raw byte data of a model resident in HBM or in motion across the network. CloakLM tackles these adversary by break assumptions around memory layout and transfers. Confidential computing mitigates ③ with latency overheads and requires the enclave to be extended for the entire boundary of trust.

② **Non-participating GPU on the same accelerator fabric.** A GPU belonging to a different tenant on the same NVLink domain or accelerator fabric represents a distinct attacker position from the host-level adversary above [25]. Operating at the device level, it has potential visibility into peer GPU memory through fabric-level access mechanisms, without requiring host-side code execution or management network reachability. This arises naturally in multi-tenant accelerator clusters where GPUs from different workloads share the same NVLink based backend network infrastructure.

③ **Physical hardware access.** An attacker with physical access to the server or datacenter infrastructure can instrument the hardware directly—for example, by inserting a PCIe snooping device into the interconnect to intercept DMA transfers between host and GPU, or by attaching instrumentation to the accelerator to read HBM contents without involving the host software stack. Unlike the VM- or driver-level adversaries above, this attacker operates below the software layer entirely. Hermes [39] demonstrates the severity of the interconnect vector, achieving lossless DNN model reconstruction by passively analyzing PCIe packets alone, without any interaction with the GPU or inference stack.

All three attacker positions share the following capability: they obtain raw memory contents or raw bus transfers from outside the serving stack. They see physical bytes, either captured from the PCIe bus or read directly from HBM, without access to the framework-level metadata that would otherwise be needed to interpret them: the PyTorch virtual memory layout, the virtual-to-physical page mapping, and the placement decisions that record which tensor occupies which address range. In an unprotected deployment, this limitation is immaterial due to the predictable layout of current ML infrastructure. As established in Section 2, ML frameworks allocate model weights in large, contiguous physical regions that are persistently resident and repeatedly accessed in a predictable order. The regularity of this layout means that raw physical bytes alone are sufficient to infer layer boundaries, tensor shapes, and parameter values, the framework metadata is effectively recoverable from the data itself [36, 39].

### 3.2 Limitations of Existing Defenses

**Enclave based mechanisms leave most inference deployments exposed.** CC and TEE offer hardware enclave but impose non-trivial latency overhead (~10-20% [2, 29]), limiting deployment in latency-sensitive production settings. As a result, the majority of inference today runs without CC protection, leaving interconnect transfers and HBM contents exposed to the attack surfaces described above. Where CC is deployed, it protects data in motion and enforces access control over HBM, but does not encrypt HBM contents at rest, a potential residual exposure for adversaries with physical device access below the access-control layer.

**The CC perimeter is bounded by vendors's interconnect domain.** CC enclave boundary encompasses the CPU–GPU PCIe link and NVLink connections within a domain. It does not extend to the broader network fabric. In multi-rack or heterogeneous deployments, inter-rack communication traverses InfiniBand or Ethernet without CC-provided encryption. NVIDIA Vera Rubin NVL72 [23] is the largest CC-enabled protected domain with a full rack-scale NVLink fabric, but multi-rack and heterogeneous accelerator deployments remain outside this boundary. A single non-CC-capable node in a distributed inference cluster is sufficient to break the end-to-end isolation guarantee. CC and CloakLM therefore address complementary threat classes: CC enforces who may access the accelerator and encrypts data in motion, while CloakLM degrades the utility of raw memory contents for adversaries who operate at or below the access-control boundary.

### 3.3 Threat Model

We consider an adversary operating in the same physical datacenter as a deployed inference service, occupying one or more of the following positions: (1) physical or firmware-level access to the PCIe bus or accelerator interconnect, enabling passive observation of DMA transfers; (2) a non-participating or compromised host on the same infrastructure, enabling bus-level reads of HBM contents via under-utilization exploits or (3) non participating GPUs on the same accelerator fabric, allowing NVLink or RDMA based unauthorized memory accesses. Figure 2 illustrates these positions.

We assume that the host executing the inference workload, including the CPU, operating system, hypervisor, GPU drivers, and software stack, is trusted and uncompromised. The adversary operates outside the serving stack, leveraging physical or bus-level access to observe data transfers or extract device memory contents. We explicitly exclude attacks that rely on host-side compromise, software privilege escalation, or malicious kernel execution.

The security goal of CloakLM is not to make weight extraction impossible: an adversary with unrestricted physical control of a device can always recover data at some level. Rather, CloakLM aims to *impose cost*, to ensure that raw physical bytes, whether captured from the bus or dumped from HBM, do not yield an interpretable, deployable model without effort commensurate with training from scratch. We treat reconstruction cost as the operative security metric: CloakLM succeeds if the effort required to recover a functionally accurate model from raw observations exceeds the effort of training an equivalent model independently.

### 3.4 Model Exfiltration Attacks

**Model extraction via PCIe snooping with Hermes [39].** An adversary may insert a PCIe snooping device into the interconnect to intercept packets exchanged between the host and the GPU. Prior work demonstrates that PCIe packets expose physical addresses corresponding to transferred data [39]. When combined with the observation that GPU memory allocations for model parameters are typically large and contiguous, these traces enable accurate reconstruction of both model architecture and weights.

Leveraging this property, along with the assumption that GPU memory allocations for model layers and parameters are largely contiguous, Hermes trains reconstruction models that can recover architectures and weights with high accuracy, even for previously unseen models. These techniques further assume that layers are transferred to the GPU in the same order in which they appear in the model graph.

**Model extraction via GPU device memory access with TunnelS [36].** Beyond interconnect observation, an adversary with physical access to the accelerator can directly extract GPU high-bandwidth memory (HBM) contents without interfering with normal execution. Such attacks operate via DMA-based reads over the PCIe fabric or external instrumentation of the accelerator memory interface, achieved by modifying device drivers, exploiting covert memory-access side channels, or introducing malicious OS APIs, without modifying the user-facing host software stack or executing malicious GPU kernels, and are therefore consistent with our trusted-host assumption. The resulting memory dumps expose all data resident in GPU memory, including model parameters and activations. As with PCIe-based attacks, existing reconstruction techniques rely on frameworks allocating parameters in large, contiguous HBM regions, preserving the semantic structure needed to identify and recover individual layers and tensors.

## 4 CloakLM

### 4.1 Memory Management in Serving

LLM serving frameworks manage GPU memory through a multi-tiered hierarchy: vLLM orchestrates resource allocation, PyTorch handles device-level memory management, and the GPU runtime maps virtual to physical pages in HBM. This hierarchy determines where model weights and KV caches reside, how long they persist, and which surfaces are visible to an accelerator-level adversary. Figure 3 illustrates this stack.

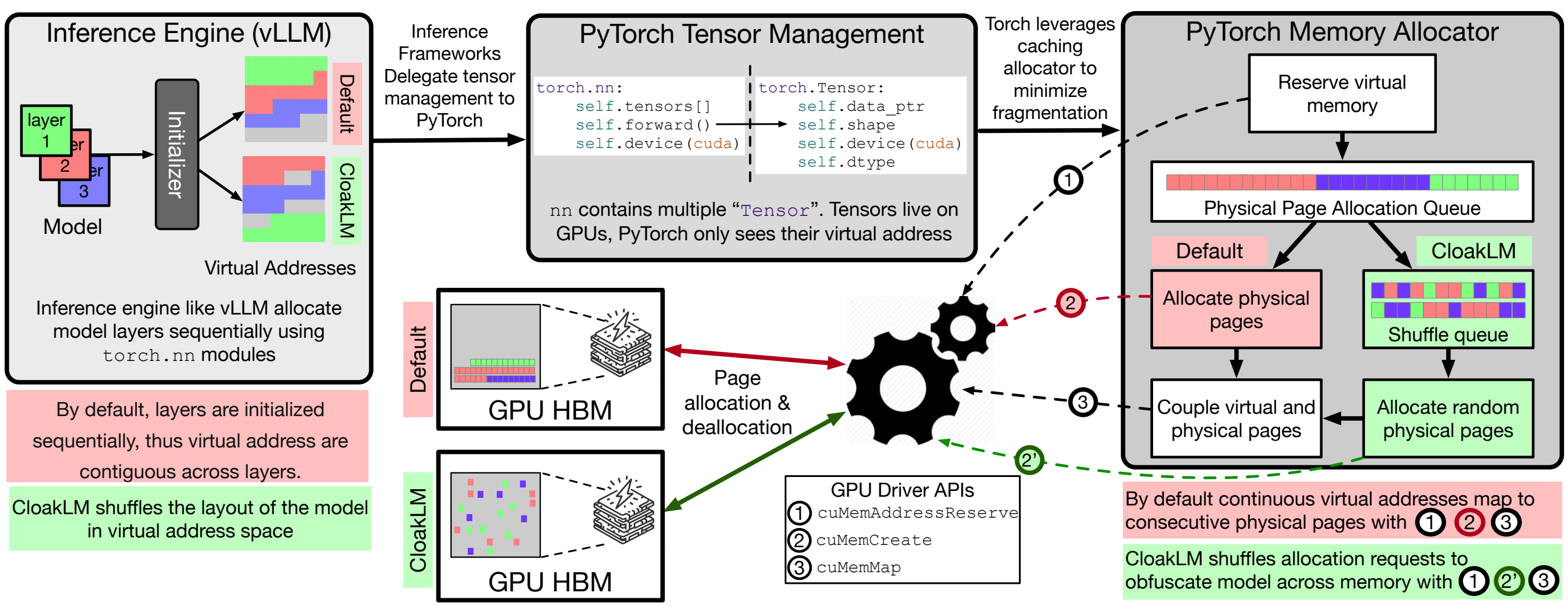

**Figure 3.** LLM frameworks such as vLLM use PyTorch to allocate and manage logical tensors via a CUDA caching allocator. By default, PyTorch reserves large contiguous regions of GPU virtual memory and maps it to contiguous physical pages, creating predictable patterns exploitable by memory-based attacks. In contrast, CloakLM introduces inter-layer and page-level obfuscation, randomizing both virtual-to-physical mappings and tensor layout to disrupt structural regularity while preserving inference correctness.

**User Frontend: vLLM.** Frameworks such as vLLM expose a high-level interface that allows users to configure and instantiate LLMs with minimal effort (e.g., `llm = vllm.LLMEngine (...)`). Internally vLLM orchestrates a complex sequence of memory allocations and data movements, including loading model weights, managing key–value (KV) cache blocks for auto-regressive decoding, and tracking active tensors. vLLM delegates managing device memory and these low-level responsibilities to the PyTorch backend.

**Developer Frontend: PyTorch.** PyTorch further abstracts GPU memory management by exposing tensor-based APIs that allow developers to allocate device memory using simple, declarative commands (e.g., `tensor = torch.tensor(size, device='cuda')`) Internally, PyTorch employs a caching allocator designed to minimize the frequency of allocation and deallocation requests issued to the GPU over the PCIe. To achieve this, PyTorch's caching allocator exploits the separation between virtual and physical address spaces. It reserves large contiguous regions of virtual address space upfront while mapping physical HBM pages on demand. While this design improves performance, it also results in long-lived physical pages that may be reused across tensors, requests, or users—an important consideration under an accelerator-level adversary.

**Leveraging Virtual–Physical Address Decoupling.** The decoupling of virtual and physical address spaces provides significant flexibility in how data is placed and managed within GPU memory. Applications operate exclusively on virtual addresses, while the runtime and device driver transparently manage the underlying physical page mappings. Consequently, physical memory layouts can be modified, such as rearranging, remapping, or shuffling pages, without affecting application correctness or developer-visible semantics. This indirection creates an opportunity to introduce defenses that disrupt an adversary's view of GPU-resident data while preserving the abstraction expected by existing LLM serving frameworks. This property forms a key enabling mechanism for CloakLM.

### 4.2 CloakLM's Defense

CloakLM's goal is to ensure that raw memory observations, whether captured from the PCIe bus or extracted from HBM, do not yield a reconstructible model by introducing structured memory obfuscation during model loading and execution. The memory layout properties that make inference weights an attractive target, stable residency, contiguous allocation, predictable access order, also make them amenable to software-level obfuscation. CloakLM is designed to degrade the effectiveness and accuracy of model ex-filtration, extraction, and reconstruction attacks while preserving inference correctness and imposing negligible performance overhead.

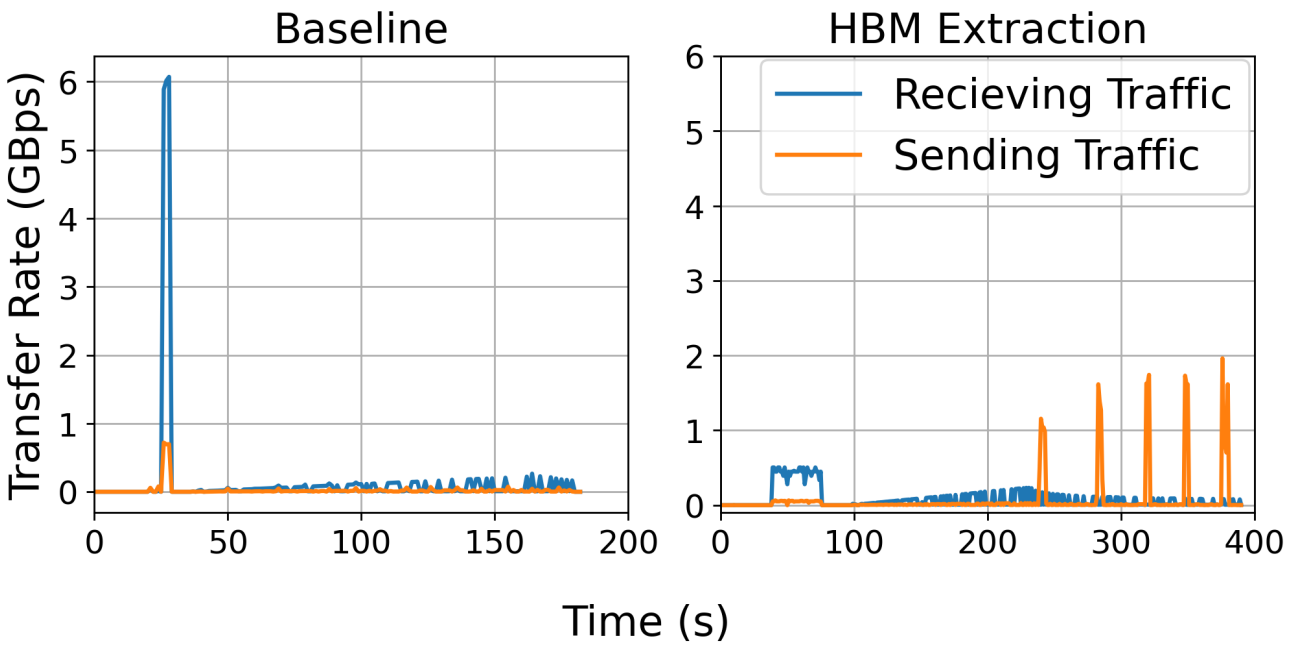

**Figure 4.** PCIe utilization for inference. *Left:* Bus utilization is relatively low, with the maximum utilization happening when the model weights are transferred during initialization. The peak of 6GBps observed in this case is much lower than the maximum bandwidth of the link, 32GBps. *Right:* Prior model extraction attacks have taken advantage of this free resource to extract the model quickly [36].

An adversary reconstructing weights from bus intercepts or HBM dumps relies on three assumptions: that layers appear in model order, that tensors occupy contiguous physical regions, and that layout persists long enough to correlate observations across time. CloakLM targets each assumption directly, operating entirely within the serving software stack without relying on cryptographic isolation or trusted hardware. It applies obfuscation at three complementary levels of the memory hierarchy:

1. **Interconnect congestion:** Saturates idle PCIe bandwidth to raise the cost and reduce the signal of bus-level observation.
2. **Inter-layer virtual address shuffling:** Randomizes the order in which layers are initialized and mapped, decoupling model structure from observable memory layout.
3. **Physical page remapping:** At initialization and then periodically randomizes the mapping of virtual addresses to physical HBM pages, breaking spatial contiguity and temporal consistency.

Figure 3 illustrates how CloakLM modifies the default memory allocation of vLLM and PyTorch across these three levels.

**Interconnect obfuscation via PCIe congestion.** Inference transfers model weights to HBM once at initialization and accesses them in place thereafter. As shown in Figure 4, PCIe utilization during steady-state inference remains well below link capacity, a window that attacks such as TunnelS exploit to exfiltrate HBM contents with minimal contention. CloakLM closes this window by issuing parallel, benign read and write operations on the PCIe interconnect during inference, inflating observable traffic, reducing the signal-to-noise ratio available to a snooping adversary, and introducing contention that directly slows memory dump throughput.

Production LLM inference at scale typically occupies dedicated nodes, as model sizes and memory requirements preclude node-level co-tenancy for most serving configurations. In shared deployments where co-tenants exist, their traffic independently adds noise to the observable bus. Since PCIe bandwidth is a local bus resource shared only among co-located tenants, the additional traffic issued by CloakLM occupies only the idle headroom already available on the link. Because inference places modest demands on PCIe bandwidth, this additional traffic does not measurably affect inference latency or throughput. Even in deployments where interconnect congestion is disabled, the remaining two defense tiers, virtual address shuffling and physical page remapping, described below, independently disrupt weight reconstruction from raw bus or HBM observations.

**Inter-layer randomization via virtual address shuffling.** PCIe-based extraction attacks assume that model layers are transferred to the GPU in architectural order and occupy sequentially increasing virtual addresses. CloakLM violates both assumptions by randomizing the order in which layers are instantiated and mapped during initialization, decoupling logical layer index from virtual address placement. By violating these assumptions, inter-layer randomization obscures the correspondence between observed PCIe transfers or virtual memory regions and the logical model structure. As a result, attackers are forced to solve a substantially harder reconstruction problem without a reliable notion of layer boundaries or ordering.

As shown in Table 1a, default allocators produce monotonically increasing addresses that directly expose model order. An attacker observing PCIe traffic under this layout can infer that adjacent address regions correspond to consecutive layers and estimate layer sizes from address extents. In contrast, CloakLM permutes the order in which tensors are transferred and mapped, decoupling virtual address ordering from model structure. Adjacent virtual regions no longer imply adjacency in the model graph, nor do address ranges reveal layer sequencing.

This shuffling is applied once at initialization, the only phase in which the full set of weight transfers crosses the PCIe bus. After initialization, weights remain resident in HBM and inference relies on a small number of stable virtual handles for kernel dispatch; re-shuffling during steady-state inference therefore provides little additional protection against interconnect-based attacks. A potential recovery vector [7] is kernel launch metadata, which could in principle reveal execution order. This signal is largely unavailable in modern LLM inference: production pipelines rely on CUDA Graphs, which capture and replay kernel sequences from a single profiling pass and expose no runtime trace of layer-level dispatch order.

**Physical page remapping.** Virtual address shuffling disrupts layer ordering but does not break the physical contiguity of individual tensors. CloakLM addresses this with

| Layer Index | Addresses | | Jumps | |
|---|---|---|---|---|
| | Default | CloakLM | Default | CloakLM |
| 1 | 0x13554 | 0xF7980 | — | — |
| 2 | 0x13584 | 0xF8280 | 0x00030 | 0x90000 |
| 3 | 0x135B4 | 0x13524 | 0x00030 | -0x3CFC0 |
| 4 | 0x135E4 | 0x507E4 | 0x00030 | 0x3D2C0 |
| 5 | 0x13614 | 0x207E4 | 0x00030 | -0x30000 |

**(a)** Virtual address layout. Default allocators produce a uniform 0x30 stride that exposes layer order; CloakLM shuffles layer placement to produce variable, non-informative jumps.

| Tensor (Page) | Page Index | | Contiguous? | |
|---|---|---|---|---|
| | Default | CloakLM | Default | CloakLM |
| T1 (P1) | 0x0027a | 0x00134 | — | — |
| T1 (P2) | 0x0027c | 0x002d4 | Yes | No |
| T2 (P1) | 0x0027e | 0x00332 | Yes | No |
| T2 (P2) | 0x00280 | 0x002fa | Yes | No |
| T2 (P3) | 0x00282 | 0x00330 | Yes | No |

**(b)** Physical page allocation. Default allocators assign sequential 2 MB pages to tensors; CloakLM remaps tensors to non-contiguous physical pages, breaking spatial layout assumptions.

**Table 1.** CloakLM disrupts the contiguous memory layout that reconstruction attacks depend on, at two levels of the memory hierarchy. Red indicates the predictable structure exploited by attackers; green indicates the obfuscated layout produced by CloakLM.

periodic randomization of the mapping between virtual address ranges and physical HBM pages using CUDA Virtual Memory Management (VMM) APIs. As shown in Table 1b, default allocators assign contiguous 2 MB pages to tensors sequentially; CloakLM remaps these to non-contiguous, non-sequential physical pages, eliminating the spatial regularity on which HBM dump reconstruction depends. From the inference engine's perspective, tensor pointers and access semantics remain unchanged, CUDA VMM makes the remapping transparent to the application. An adversary observing physical HBM instead encounters fragmented, time-varying layouts that invalidate assumptions of spatial locality and temporal persistence.

Physical page remapping introduces no additional memory footprint; The only overhead arises from synchronization points needed to safely remap pages, which can be amortized over long inference intervals or triggered selectively based on the anticipated attack model (e.g., continuous snooping versus one-shot memory dumping). Because inference workloads are dominated by repeated reuse of resident weights, these remapping operations do not affect steady-state throughput or inference latency. This exposes a tunable tradeoff: operators can adapt reshuffling frequency to their deployment risk without modifying the model or inference stack. We evaluate this tradeoff in Section 5 showing that physical page shuffling imposes negligible overhead on end-to-end inference latency when performed once every 1,000 inference iterations, relative to the scale of billions of inference requests in production serving deployments.

**Performance impact of memory obfuscation.** LLM inference issues large, streaming memory requests spanning gigabytes of weights and KV caches. Although DRAM access latency can be affected by bank conflicts and row-buffer misses, these effects are most pronounced under small, serialized accesses. LLM inference is bandwidth-bound: kernels saturate HBM bandwidth across many parallel requests rather than exploiting physical page contiguity. When CloakLM randomizes physical page placement, large accesses distribute naturally across HBM banks, maintaining uniform bandwidth utilization. Section 5 confirms empirically that physical page remapping incurs negligible impact on end-to-end inference latency.

## 5 Evaluation

We evaluate CloakLM against two concrete prior attacks, a memory dump attack based on TunnelS [36] and a PCIe snooping attack based on Hermes [39]. We evaluate our strategy across three broad metrics: (i) accuracy of model extracted, (ii) impact of extraction process, and (iii) latency impact on end-to-end inference latency.

**Setup and Hardware.** We perform our experiments on L40s GPUs with 46GB HBM memory and PCIe 4.0 connectivity to the CPU with 16 lanes (32GB/s uni-directional bandwidth). We conduct our experiments on three models: Llama 3.1 (8B parameters), Qwen 3 (14B parameters) and Qwen 3 MoE (30B parameters, 3B active parameters) for the attacks using vLLM framework, evaluate that CloakLM significantly reduces the risk of model exfiltration and has no adverse effects on the end-to-end latency of LLMs. Our set up is detailed in Table 2.

**Implementation Details.** CloakLM integrates its defence strategies directly into the PyTorch and vLLM codebases, exposing a simple one-line API for service providers in ~300 lines of code. All changes are backwards compatible across vLLM versions. Since the host is trusted, we seed the PRNG from `/dev/urandom` prior to deployment: virtual addresses are enumerated, a random permutation is computed, and physical pages are allocated accordingly, all before any weight transfer ensuring the attacker never observes a structured PCIe memory layout.

### 5.1 Attack Evaluation

**Threat assumptions.** Existing PCIe- and memory- based attacks rely on two key structural assumptions: (1) model parameters are allocated in large, contiguous memory regions and (2) the attacker can recover meaningful virtual-to-physical layout information. Under these conditions, attackers exploit layout regularities to identify layer boundaries and reconstruct parameters.

| Model Name | Model Type | Parameters #Billion | Tensor Parallelism | Virtual-Physical Contiguity | | Shuffling Latency(s) |
|---|---|---|---|---|---|---|
| | | | | Default | CloakLM | |
| Llama 3.1 | Dense | 8 | 1 | 91.41% | 2.38% | 1.58 |
| Qwen 3 | Dense | 14 | 1 | 95.57% | 1.57% | 1.61 |
| Qwen 3 | Dense | 14 | 2 | - | - | 2.81 |
| Qwen 3 | MoE | 30 | 2 | - | - | 2.95 |

**Table 2.** We evaluate 3 models across 4 configurations (dense and MoE, varying tensor parallelism), reporting the percentage of virtual address ranges backed by contiguous physical pages under the default and CloakLM allocator. CloakLM reduces virtual–physical contiguity from over 90% to below 3%, disrupting layout-based reconstruction. Since shuffling is performed independently per GPU, similar reductions hold under TP2. Shuffling latency remains modest at 1.5–3 seconds per configuration.

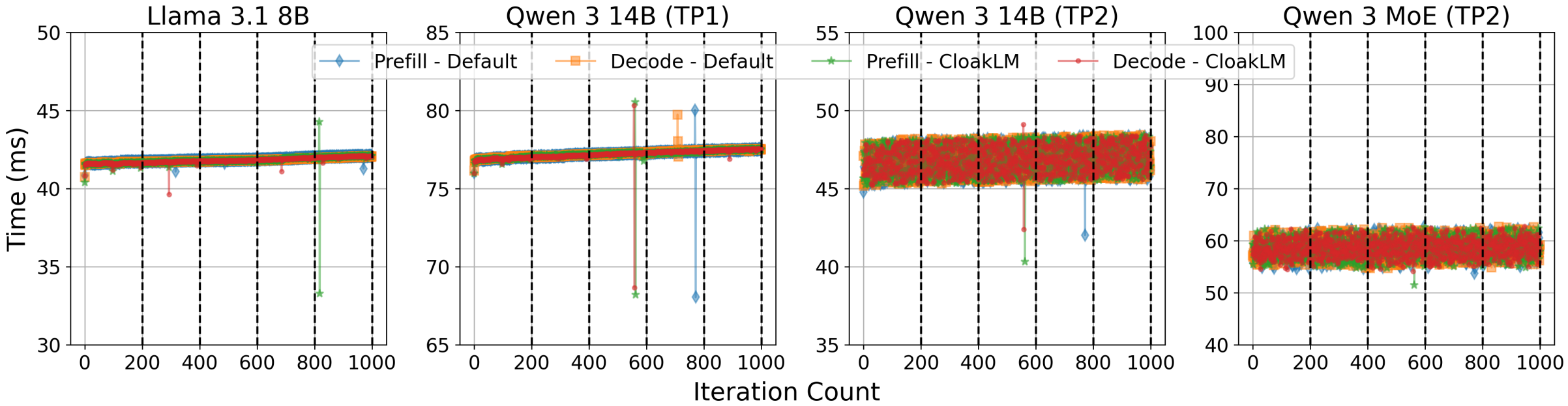

**Figure 5.** Iteration-level prefill and decode latency with and without CloakLM. Across all evaluated models and parallelism settings, CloakLM incurs no measurable runtime overhead, with latency distributions closely matching default execution.

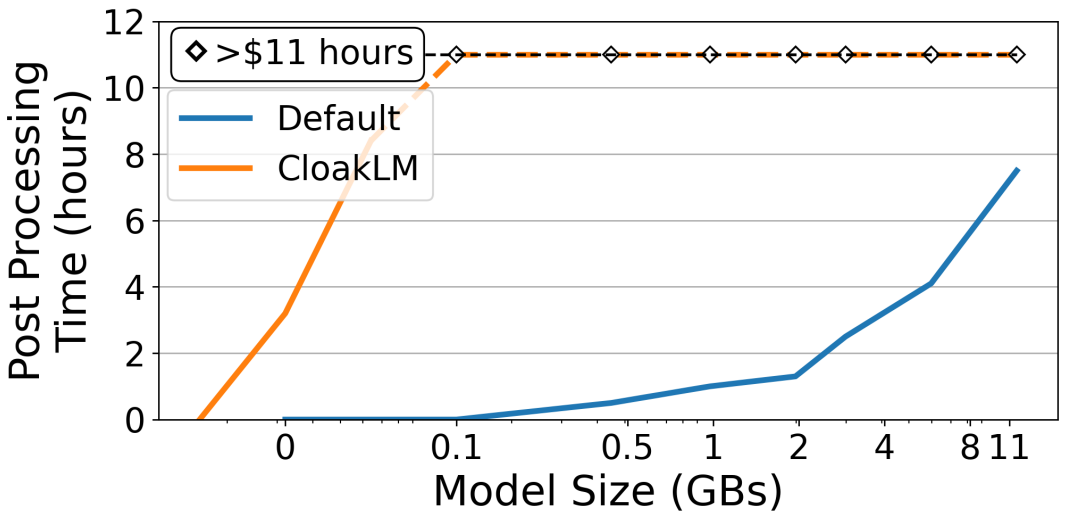

**Figure 6.** Extracting models larger than 64 MB in size is impractical with CloakLM. However, it is possible to extract models as large as 11 GBs with naive methods.

Unless stated otherwise, we assume the attacker knows the target model architecture and has access to at least one correct input–output pair for validation. This gives the adversary maximal leverage and represents a best-case setting for reconstruction. Our defense does not rely on obscuring the architecture: prior work has shown that architectures can often be inferred from memory traces, kernel launches, or I/O behavior [22, 31, 36, 39]. Our goal is to prevent reconstruction *even under strong attacker knowledge.*

**HBM dump attack based on TunnelS [36].** TunnelS demonstrates that full HBM exfiltration is achievable from a non-participating host using a modified NVIDIA driver, without access to the victim host or its workload, by exploiting idle PCIe bandwidth during inference. We obtain GPU memory dumps using the technique described in TunnelS, then attempt reconstruction by grouping physically contiguous HBM pages into segments and enumerating permutations until a candidate ordering reproduces the known correct output.

Under default vLLM, we successfully reconstruct models up to 11 GB (16 LLaMA decode layer) within approximately eight hours, as shown in Figure 6. Under CloakLM, physical page dispersion eliminates stable contiguous segments entirely, rendering reconstruction infeasible within the same time budget for models as small as 64 MB. CloakLM therefore removes the structural foundations that state-of-the-art reconstruction attacks depend on, even when the attacker has strong prior knowledge of the model.

**PCIe Snooping Attack based on Hermes [39].** Hermes [39] demonstrates that a passive adversary with physical access to the PCIe bus can reconstruct both the architecture and weights of a DNN model hosted on a GPU, without any interaction with the victim system. The adversary interposes a snooping device, such as a PCIe interposer or rogue peripheral, on the host-to-device interconnect, capturing raw DMA transfer packets as tensors (activations, weights, gradients) flow between host memory and the GPU during inference. By reverse-engineering the traffic patterns (packet sizes, transfer sequences, timing intervals) Hermes reconstructs the layer topology and ultimately recovers full model weights with lossless fidelity. The attack is entirely passive, requires no software foothold on the victim host, and leaves no detectable trace in system logs.

Fully reproducing this attack requires custom PCIe snooping hardware. We instead emulate it by (i) instrumenting the PyTorch/CUDA runtime to capture tensor transfer traces during DNN inference, recording the sequence, size, and timing of every host-to-device and device-to-host DMA transfer, and (ii) replaying these traces through a packet generation model that faithfully reconstructs the PCIe traffic stream as it would appear to a physical interposer, following the threat model described in Hermes. This emulation allows us to evaluate reconstruction fidelity and measure the effect of CloakLM's defenses without requiring physical snooping hardware.

We evaluate reconstruction fidelity under three conditions: (*i*) default serving with no defenses; (*ii*) CloakLM with virtual memory shuffling only; and (*iii*) CloakLM with both virtual shuffling and PCIe congestion. Under default serving, we reproduce the Hermes result: lossless reconstruction of a CNN-based binary classifier. Virtual memory shuffling alone defeats the attack, degrading classifier accuracy from ~0.9 to ~0.5 and collapsing the reconstructed model to near-random performance, by destroying the correspondence between observed DMA addresses and model layers. Adding PCIe congestion introduces structured noise into the transfer stream, perturbing packet timing and sizes to further degrade any residual signal available to a more adaptive adversary.

### 5.2 Performance Evaluation

**Inference latency.** As shown in Figure 5, physical memory shuffling introduces no measurable overhead to per-iteration prefill or decode latency across all evaluated models and configurations, dense, MoE, single-GPU, and multi-GPU. Figure 7 confirms that performance degradation is confined to the shuffling operation itself and does not affect steady-state execution. Moreover, since token packets during inference are small (on the order of MBs), PCIe congestion likewise introduces no measurable degradation in time-to-first-token or time-between-tokens.

**Initialization overhead.** PCIe congestion increases link utilization to approximately 30 GB/s during model loading, as shown in Figure 9. This directly slows PCIe-based extraction: the time required to exfiltrate a 48 GB model increases from approximately 45 seconds to over 8 minutes. The corresponding initialization overhead scales linearly with model size at approximately 1.1 s/GB (9.1 s for 8B, 16.1 s for 14B, 35.5 s for 30B), as shown in Figure 8. Virtual address shuffling introduces negligible initialization overhead. Physical page shuffling requires approximately 1.5 s per GPU and is performed independently on each GPU in multi-GPU deployments, as reported in Table 2.

**Shuffling Overhead in Multi-GPU Environments.** Table 2 shows that HBM shuffling incurs modest overhead, requiring approximately 1.5 seconds per GPU on average. In multi-GPU deployments, shuffling is currently performed independently on each GPU.

**Virtual and Physical Address Shuffling.** Virtual address shuffling does not affect model upload time or steady-state iteration latency. Across all evaluated models and parallelism configurations, performance remains unchanged, as confirmed by Figure 8. However, as illustrated in Table 1a, the virtual addresses corresponding to model layers are permuted, disrupting their original ordering and significantly increasing the difficulty of inferring layers from PCIe traffic. Physical memory shuffling further fragments the memory layout. As shown in Table 2, the number of contiguous virtual address ranges backed by contiguous physical pages drops substantially under CloakLM. This fragmentation directly undermines key assumptions made by prior model reconstruction attacks that rely on large contiguous memory regions to infer model structure.

**Configurable security–overhead tradeoff.** As shown in Figure 7, shuffling frequency directly governs the security–overhead tradeoff: more frequent reshuffling provides stronger protection against continuous observation but increases amortized overhead. In practice, a single shuffle at initialization is sufficient to disrupt one-shot HBM dump attacks, while periodic reshuffling can be enabled to defend against adversaries with sustained observation capability. Operators can tune reshuffling frequency to match their deployment risk profile without modifying the model or inference stack.

## 6 Discussion

### 6.1 Security Guarantees and Non-Goals

In security, as new defenses are deployed, adversaries adapt and develop increasingly sophisticated attacks. Our goal is not to claim absolute protection against model exfiltration, but to raise the cost and complexity of *practical attacks that exist today or are feasible under current deployment models*, specifically, accelerator-centric attacks that exploit visibility into interconnect traffic and device memory during inference. We acknowledge that determined adversaries may eventually develop techniques that circumvent our defenses.

We focus on inference-time threats. During inference, model weights are static, long-lived, and repeatedly accessed, making the network fabric and accelerator memory hierarchy the most exposed attack surfaces for model exfiltration.

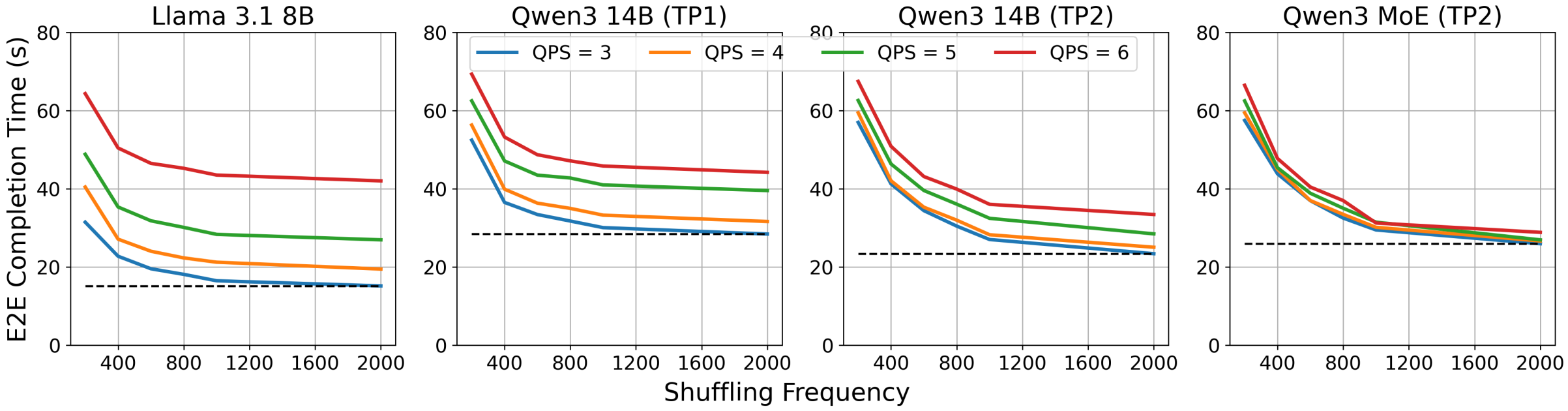

**Figure 7.** We evaluate workloads across a range of QPS values and input/output sequence lengths uniformly sampled between 100 and 200 tokens. End-to-end performance remains stable when physical shuffling is performed once every 1,500 iterations. Performance degradation increases proportionally with shuffling frequency. In practice, a single shuffle during initialization is sufficient to obfuscate HBM contents, while additional runtime shuffling can be enabled to provide stronger security guarantees.

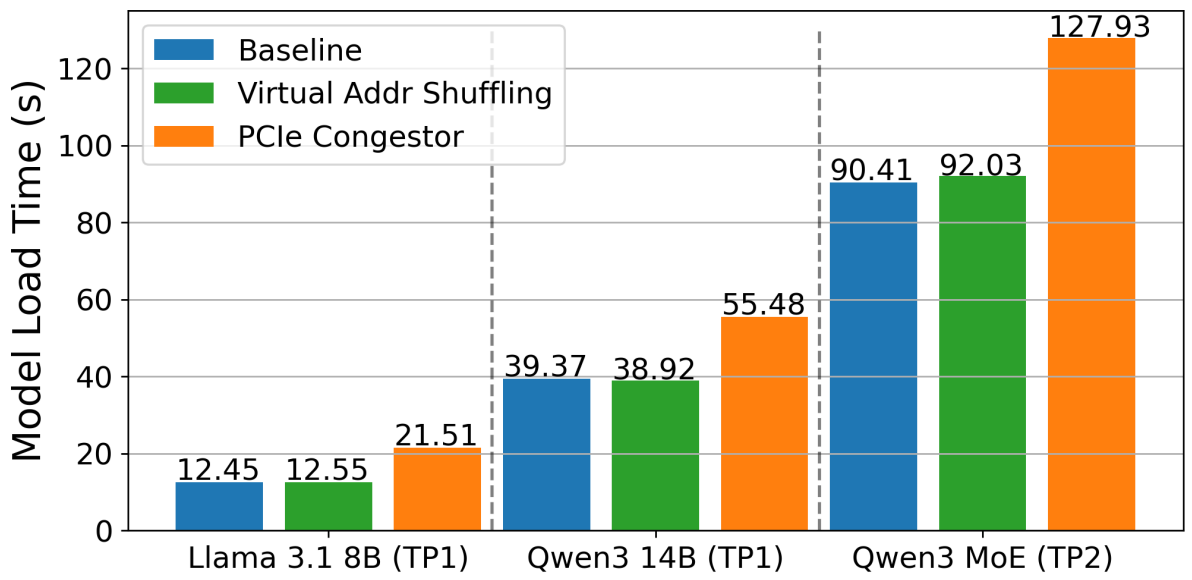

**Figure 8.** Impact of CloakLM on model initialization and loading latency. PCIe congestion increases loading time proportionally to model size, while virtual shuffling introduces negligible overhead.

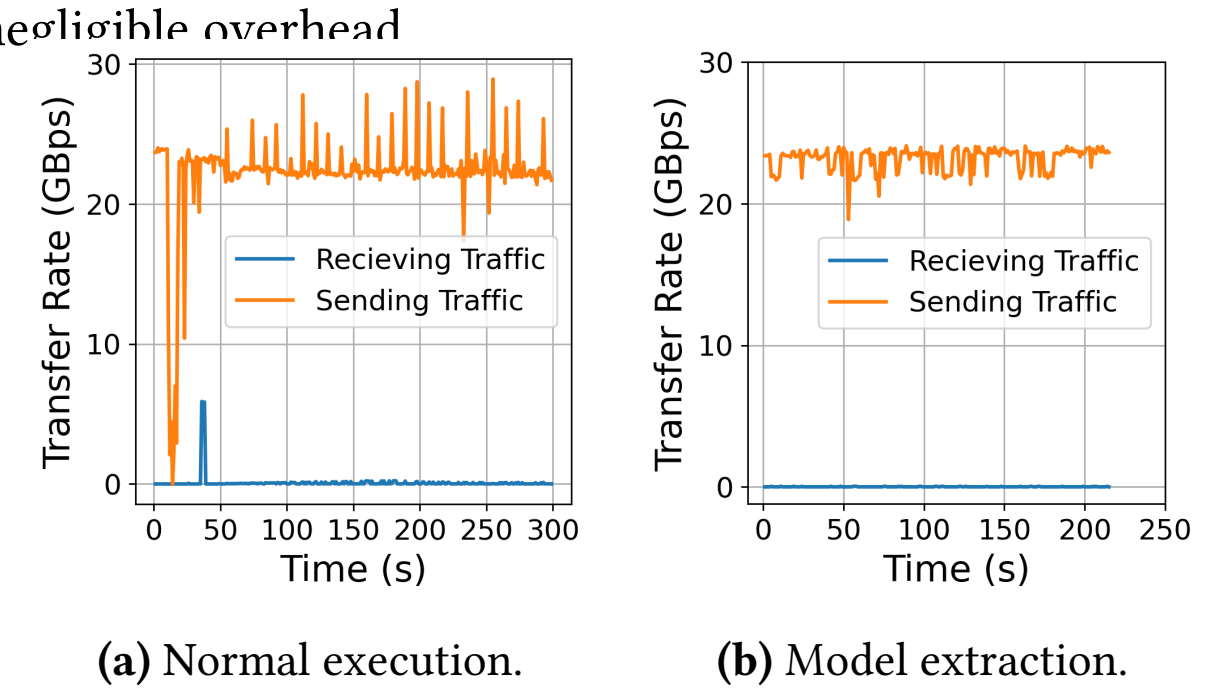

**(a)** Normal execution. **(b)** Model extraction.

**Figure 9.** PCIe traffic with interconnect congestion. Both look same and model extraction is severely delayed, from 45 seconds to extract 48GBs to over 8 minutes!

Training workloads, by contrast, continuously update parameters, rendering intermediate snapshots transient and requiring fundamentally different threat models; protecting against training-time exfiltration is left for future work.

Rather than aiming for cryptographic secrecy, CloakLM focuses on *structural obfuscation*: breaking the assumptions that existing extraction attacks rely on, such as contiguous memory layouts, stable layer ordering, and low-noise interconnect traffic, thereby degrading attack accuracy and increasing reconstruction effort. A key design constraint is that defenses must preserve inference correctness and performance, avoiding mechanisms that introduce measurable latency, reduce throughput, or require changes to model execution semantics. CloakLM thus bridges the gap between purely performance-driven systems and heavyweight confidential computing solutions such as full memory encryption or trusted accelerator enclaves.

### 6.2 Out-of-Scope Attack Vectors

The following attack vectors are orthogonal to the focus of this paper.

**Timing-based attacks.** These attacks exploit congestion and co-tenancy on shared network components to infer victim model properties [14, 28], injecting PCIe traffic and measuring latency variations across inference stack peripherals (host–device links, RDMA NICs, NVMe SSDs). Such techniques are limited to coarse properties like model size or architecture, making them substantially weaker than the weight-recovering attacks discussed above. Timing side channels over NVLink have also been explored [6, 35].

**Residual memory and cache attacks.** These attacks exploit residual memory state or shared caches under co-tenancy, relying on insufficient memory isolation or incomplete hardware state clearing in public cloud environments [12, 27]. While powerful, they assume logical co-location rather than physical access to interconnects or device memory.

**Other side-channel attacks.** Additional side channels include power consumption [8], GPU context-switch behavior [31], and application-level observables [22], typically leaking indirect information under strong assumptions about workload scheduling or observability.

**Participating host-side attacks**. CPU-based attacks spanning memory [21], cache [34, 37], and microarchitectural channels [38] are extensively studied, with well-established defenses [9, 18]. We explicitly assume the host executing inference is secure.

## 7 Conclusions

This paper identifies inference-time model exfiltration as a critical and underexplored security risk in accelerator-centric AI deployments. As model providers increasingly rely on third-party and disaggregated infrastructure, model weights are exposed to interconnect- and memory-level attacks that bypass traditional software isolation.

We present CloakLM, a lightweight, performance-preserving defense framework that disrupts the structural assumptions exploited by existing extraction attacks. By combining interconnect congestion, randomized layer initialization, and dynamic GPU page remapping, CloakLM obfuscates memory layouts and data movement patterns without modifying inference kernels or degrading throughput. Our results show that these techniques substantially raise the cost and complexity of model reconstruction while maintaining production-grade performance. We hope CloakLM motivates further exploration of systems-level defenses for model IP protection in shared AI infrastructure.